\documentclass[12pt]{article}

\begin{document}
\begin{center}
{\bf EXACT SOLUTIONS OF THE EQUATION FOR COMPOSITE SCALAR
PARTICLES IN QUANTIZED ELECTROMAGNETIC WAVES}

\vspace{5mm}
 S.I. Kruglov \footnote{E-mail:
krouglov@sprint.ca}\\

\vspace{5mm}
\textit{International Educational Centre, 2727 Steeles Ave. W, Suite 202, \\
Toronto, Ontario, Canada M3J 3G9}
\end{center}

\begin{abstract}
The equation is considered for a composite scalar particle with
polarizabilities in an external quantized electromagnetic plane
wave. This equation is reduced to a system of equations for
infinite number of interacting oscillators. After diagonalization,
we come to equations for free oscillators. As a result, exact
solutions of the equation for a particle are found in a plane
quantized electromagnetic wave of the arbitrary polarization. As a
particular case, the solution for monochromatic electromagnetic
waves is considered. The relativistic coherent states of a
particle are constructed in the case of the Poisson distribution
of photon numbers. In the limit when the average photon number
$<n>$ and the volume $V$ of the quantization trend to infinity
(but the photon density $<n>/V$ remains constant), the
wavefunction converts to the solution corresponding to the
external classical electromagnetic wave.
\end{abstract}

\section{Introduction}

In this work we find and investigate exact solutions of the
equation for composite scalar particles (for example pions, kaons
or charged composite Higgs bosons $H^{\pm }$) which possess
electromagnetic polarizabilities in the external quantized plane
electromagnetic wave of arbitrary polarization. It should be
noted, however, that the notion of an external electromagnetic
field in the relativistic theory is limited. In [1-3] some exact
solutions were considered for a scalar particle in the field of
the static uniform electric, magnetic fields, the monochromatic
classical and quantized electromagnetic wave. It is important to
study the general case of an interaction of charged composite
scalar particles with strong non-monochromatic laser waves with an
arbitrary polarization. For point-like scalar and spinor particles
such solutions were derived in [4-6]. In our case we can
investigate the effects which are connected with the interaction
of external electromagnetic waves with charges distributed inside
of a composite particles. Considering the quantized external
electromagnetic waves allow us to take into account the
interaction of a particle and photons with arbitrary frequencies
and polarizations. Here we describe a particle and electromagnetic
field quantum-mechanically. This is the most general and important
case which can be compared with the semiclassical approaches. We
interpret the complex, which consists of a particle and some
number of photons as a quasi-particle. This is an analog of the
state of an electron in the field of the crystal lattice. The
solutions obtained can be used in different calculations of
probabilities of composite scalar particle scattering processes in
the field of quantized electromagnetic waves. We notice that some
processes, however, can not be considered in the framework of
one-particle theory.

Section 2 contains the discussion of the conserved quantum numbers
for an arbitrary external quantized electromagnetic wave. The
solution of the equation, momentum and mass for a particle in the
field of a classical plane electromagnetic wave are considered in
section 3. In Sec. 4 we find wavefunction of a composite particle
in a quantized plane electromagnetic wave with arbitrary
polarization. The momentum and mass of a particle in a plane
quantized electromagnetic wave is studied in Sec. 5. Section 6
contains the solution of the equation for the particular case of
monochromatic plane waves. In Sec. 7 the coherent states of a
particle are investigated when the average number of photons
$\langle n\rangle $ and quantization volume $V$ go to infinity. In
this limit the wavefunction transforms to the solution of the
equation for a particle in the external classical electromagnetic
wave. Sec. 8 contains a conclusion.

We use units in which $\hbar =c=1$.

\section{An arbitrary external quantized electromagnetic wave}

The Lagrangian of a composite scalar particle interacting with
electromagnetic field is given by [1]
\begin{equation}
{\cal L}=-(D_\mu ^+ \varphi^*)(D_\nu \varphi ) \left(
\delta_{\mu\nu} - K_{\mu \nu }\right)-m_{eff}^2\varphi^*\varphi ,
\label{1}
\end{equation}
where $m$ is the rest mass and $m_{eff}^2=m^2\left( 1-\beta F_{\mu
\nu }^2/\left( 2m\right) \right) $ is the squared effective mass
of a particle, $ D_\mu =\partial _\mu -ieA_\mu $, $ D_\mu^+
=\partial _\mu +ieA_\mu $, $\partial _\mu =\partial /\partial
x_\mu $, $x_\mu =(\mathbf{x},ix_0)$, $x_0\equiv t$ is the time,
$e$ is a charge, $ A_\mu $ is the vector potential of the external
electromagnetic field, $ K_{\mu \nu }=\left( \alpha +\beta \right)
F_{\mu \alpha }F_{\nu \alpha }/m$, $F_{\mu \nu }=\partial _\mu
A_\nu -\partial _\nu A_\mu $ is the strength tensor, and $\alpha
$, $\beta $ are electric and magnetic polarizabilities of a
particle, respectively. From the Lagrangian (1) we find the
equation for the wavefunction of a composite scalar particle
\begin{equation}
D_\mu ^2\varphi -D_\mu \left[ K_{\mu \nu }\left( D_\nu \varphi
\right) \right] -m_{eff}^2\varphi =0 . \label{2}
\end{equation}
Equation (2) is a generalization of a Klein-Gordon equation on the
case of a composite scalar particle. This equation was obtained
[7] using the instanton vacuum model (that is QCD-motivated) and
the phenomenological approach [8].

Using the general expression for the density of the electric
current
\[
j_\mu = -i\left(\frac{\partial{\cal L}}{\partial(\partial_\mu
\varphi)}\varphi-\frac{\partial{\cal L}}{\partial(\partial_\mu
\varphi^*)}\varphi^*\right) ,
\]
we find from Eq. (1)
\begin{equation}
j_\mu=i(\varphi D_\nu ^+ \varphi^*-\varphi^*D_\nu \varphi)\left(
\delta_{\mu\nu} - K_{\mu \nu }\right) .\label{3}
\end{equation}
The last term in Eq. (3) contributes to the usual current density
of a point-like scalar particle. This contribution is due to the
charge distribution inside of a composite scalar particle. The
vector potential of the quantized electromagnetic field can be
chosen in the form [9]
\begin{equation}
A_\mu =\sum_{\mathbf{k},s}\frac{e_{s\mu }}{\sqrt{2k_0V}}\left[
c_{ks}^{-}e^{i(kx)}+c_{ks}^{+}e^{-i(kx)}\right] , \label{4}
\end{equation}
where $e_{s\mu }=\delta _{s\mu }$ are two unit polarization
vectors ($s=1,2$ ), $k_\mu =(\mathbf{k},ik_0)$ is a wavevector
with the properties $(e_1k)=\left( e_2k\right)
=k^2=\mathbf{k}^2-k_0^2=0$ and $k_0$ is the photon energy, $V=L^3$
is the normalizing volume ($L$ is the normalizing length). The
creation $c_{ks}^{+}$ and annihilation $c_{ks}^{-}$ operators
satisfy the commutation relations $[c_{ks}^{-},c_{k^{\prime
}s^{\prime }}^{+}]=\delta _{kk^{\prime }}\delta _{ss^{\prime }}$.
The scalar product of four-vectors is defined as $\left( kx\right)
=\mathbf{kx}-k_0x_0$. We consider in (4) an arbitrary number of
modes $N$ of electromagnetic waves.

As noted in [10], the energy-momentum tensor of a hole system is
the sum of tensors of energy-momentum of electromagnetic fields
and particles, and the last are considered as non-interacting
particles. Thus, we introduce the operator of the total
four-momentum of the particle-photon system as follows
\begin{equation}
\hat{Q}_\mu =-i\partial _\mu +\sum_{\mathbf{k},s}k_\mu \left(
N_{ks}+\frac 12\right) , \hspace{0.5in}N_{ks}=c_{ks}^{+}c_{ks}^{-}
, \label{5}
\end{equation}
where $N_{ks}$ is the operator of the photon number corresponding
to the four-momentum $k_\mu $ and polarization vector $e_{s\mu }$.
As the operator $ \hat{Q}_\mu $ commutes with the covariant
derivative $D_\mu $, i.e. $[\hat{Q}_\mu ,D_\nu ]=0$, an eigenvalue
of the operator $\hat{Q}_\mu $ is the integral of motion. As a
result, the wavefunction of a particle in the external quantized
electromagnetic field, $\varphi $, obeys as Eq. (2) and the
following equation
\begin{equation}
\hat{Q}_\mu \varphi =q_\mu \varphi ,  \label{6}
\end{equation}
where $q_\mu $ is an eigenvalue of the operator $\hat{Q}_\mu $. We
will look for the solution of Eqs. (2), (6) which corresponds to
the definite four-momentum $q_\mu $ of a particle-photon system.
It is convenient to choose the replacement as follows
\begin{equation}
\varphi =U_1\Phi ,\hspace{0.5in}U_1=\exp \left\{
-i\sum_{\mathbf{k},s}(k_\mu x_\mu )\left( N_{ks}+\frac 12\right)
\right\} .  \label{7}
\end{equation}

The operator $U_1$ is the unitary operator so that $U_1^{+}U_1=1$.
From Eqs. (5)-(7) we find
\begin{equation}
-i\partial _\mu \Phi =q_\mu \Phi .  \label{8}
\end{equation}
The solution to Eq. (8) is given by
\begin{equation}
\Phi =\exp \{i(q_\mu x_\mu )\}\chi , \label{9}
\end{equation}
where the function $\chi $ does not depend on the coordinates
$x_\mu $ and depends only on photon variables. As a result the
general solution to Eqs. (2) and (6) can be written as
\begin{equation}
\varphi =U\chi ,\hspace{0.5in}U=\exp \{i(q_\mu x_\mu )\}U_1 .
\label{10}
\end{equation}

The function $\chi $ obeys the equation
\begin{equation}
U^{+}D_\mu ^2U\chi -U^{+}D_\mu \left[ K_{\mu \nu }\left( D_\nu
U\chi \right) \right] -m^2\chi =0 , \label{11}
\end{equation}
where $U^{+}$ is the Hermitian conjugated operator:
\begin{equation}
U^{+}=\exp \left\{ -i\left[ (q_\mu x_\mu
)-\sum_{\mathbf{k},s}(k_\mu x_\mu )\left( N_{ks}+\frac 12\right)
\right] \right\} .  \label{12}
\end{equation}

We took into consideration that for electromagnetic waves with the
vector potential (4) $F_{\mu \nu }^2=0$ (and therefore
$m_{eff}=m$).

Using the operator identity
\begin{equation}
U^{+}c_{ks}^{\pm }U=\exp \left\{ \pm (k_\mu x_\mu )\right\}
c_{ks}^{\pm } ,\label{13}
\end{equation}
we find
\begin{equation}
U^{+}D_\mu U=i\left\{ q_\mu -\sum_{\mathbf{k},s}k_\mu \left(
N_{ks}+\frac 12\right) -e\sum_{\mathbf{k},s}\frac{e_{s\mu
}}{\sqrt{2k_0V}}\left[ c_{ks}^{-}+c_{ks}^{+}\right] \right\} ,
\label{14}
\end{equation}
\begin{equation}
U^{+}F_{\mu \nu }U=i\sum_{\mathbf{k},s}\frac{k_\mu e_{s\nu }-k_\nu
e_{s\mu } }{\sqrt{2k_0V}}\left[ c_{ks}^{-}-c_{ks}^{+}\right] .
\label{15}
\end{equation}

It is convenient to use the coordinate representation for the
operators $ c_{ks}^{-}$, $c_{ks}^{+}$ [9]:
\begin{equation}
c_{ks}^{-}=\frac 1{\sqrt{2}}\left( \xi _{ks}+\frac \partial
{\partial \xi _{ks}}\right) ,\hspace{0.5in}c_{ks}^{+}=\frac
1{\sqrt{2}}\left( \xi _{ks}-\frac \partial {\partial \xi
_{ks}}\right) . \label{16}
\end{equation}

In this representation the operator of the photon number becomes

\begin{equation}
N_{ks}=\frac 12\left( \xi _{ks}^2-\frac{\partial ^2}{\partial \xi
_{ks}^2} -1\right) .  \label{17}
\end{equation}

Then the wavefunction $\chi $, which obeys Eq. (11), depends on
photon variables $\xi _{ks}$.

\section{A classical plane electromagnetic wave}

Equation (2) has exact solutions for a particle moving in the
classical plane electromagnetic wave [1]. The vector potential of
a plane electromagnetic wave depends only on $\vartheta=k_\mu
x_\mu$, i.e. $A_\mu=A_\mu (\vartheta)$, so that the Lorentz
condition $\partial_\mu A_\mu =0$ is valid. For such a potential
Eq. (2) takes the form
\begin{equation}
\left(\partial_\mu ^2 -2ieA_\mu \partial_\mu - e^2 A_\mu^2 -
Wk_\mu k_\nu (A_\alpha')^2\partial_\mu \partial_\nu
-m^2\right)\varphi (x)=0 , \label{18}
\end{equation}
where $W=(\alpha +\beta )/m$, $A_\alpha'=\partial A_\alpha
/\partial \vartheta$. The solution to Eq. (18) can be represented
as
\begin{equation}
\varphi (x)=\exp (ipx)\chi(\vartheta) , \label{19}
\end{equation}
where $px=p_\mu x_\mu$, $p_\mu$ is the momentum of a free particle
so that $p^2=-m^2$. Using Eq. (19) we arrive from Eq. (18) at
\begin{equation}
2i(pk)\chi'(\vartheta)+\left[2e(Ap) - e^2 A^2 + W(kp)^2 (A')^2
\right]\chi (x)=0 . \label{20}
\end{equation}
After integration of Eq. (20), the wavefunction (19) takes the
form
\begin{equation}
\varphi (x)=C\exp
\left\{i(px)+i\int_0^{kx}\left[\frac{e(Ap)}{(kp)} - \frac{e^2
A^2}{2(kp)} + \frac12 W(kp)(A')^2 \right]d\vartheta \right\} .
\label{21}
\end{equation}
The normalization constant $C$ can be chosen as $C=1/\sqrt{2}$.
From Eq. (3) taking into account the solution (21) we find the
current density of a composite scalar particle
\begin{equation}
j_\mu=p_\mu-eA_\mu+k_\mu\left[\frac{e(Ap)}{(kp)} - \frac{e^2
A^2}{2(kp)} - \frac12 W(kp)(A')^2 \right] . \label{22}
\end{equation}
The last term in Eq. (22) came from the charge distribution inside
of a composite scalar particle. For point-like particle $W=0$, and
this term is absent.

Using solution (21), it is easy to find also the density of
kinetic momentum of a particle $P_\mu$ in the state $\varphi$.
Taking into consideration the operator of kinetic momentum [11]
$\hat{P}=-i\partial_\mu-eA_\mu$, we find
\[
P_\mu\equiv\varphi^*\left(-i\partial_\mu-eA_\mu\right)\varphi-\varphi\left(
i\partial_\mu+eA_\mu\right)\varphi^*=
\]
\begin{equation}
p_\mu-eA_\mu+k_\mu\left[\frac{e(Ap)}{(kp)} - \frac{e^2 A^2}{2(kp)}
+ \frac12 W(kp)(A')^2 \right] . \label{23}
\end{equation}
If $A_\mu (\vartheta)$ is a periodic function, the mean value of
$A_\mu (\vartheta)$ equals zero, $\overline{A}_\mu (\vartheta)=0$.
Then averaging Eq. (23) on time, we arrive at the mean value of
the kinetic momentum density of a composite particle
\begin{equation}
\overline{P}_\mu=p_\mu-k_\mu\left[\frac{e^2 \overline{A^2}}{2(kp)}
- \frac12 W(kp)\overline{(A')^2} \right] . \label{24}
\end{equation}
From Eq. (24) we find
\begin{equation}
\overline{P}_\mu^2=-m_*^2,\hspace{0.3in} m_*^2=m^2+e^2
\overline{A^2} - W(kp)^2\overline{(A')^2} , \label{25}
\end{equation}
where $m_*$ is the effective mass of a composite particle in the
field of a classical electromagnetic wave. When we neglect the
polarizabilities of a particle, i.e. $W=0$, we come to the known
expression [11]. The total momentum of the particle-photon system
(see Eq. (5)) can be represented as follows
\begin{equation}
q_\mu =P_\mu +R_\mu ,
 \label{26}
\end{equation}
where $R_\mu$ is the four-momentum of photons (the electromagnetic
field) interacting with a particle. In the framework of
one-particle theory both variables $P_\mu$, $R_\mu$ are the
integrals of motion (see also [6]).

\section{Plane quantized electromagnetic waves}

For our purposes, we will use the approximation of a plane wave
when the momenta of photons are parallel to the direction $\kappa
_\mu =(\mathbf{\kappa } ,i\kappa _0)$ so that $\mathbf{\kappa
}^2=\kappa _0^2$, $\kappa _0=2\pi /L$ and $k_\mu =n\kappa _\mu $,
where $n$ is an integer positive number $ n=1,2,...,N$. All
frequencies of photons are taken into consideration here. In the
case of the monochromatic wave only one term survives in the sum
of Eq. (4). For a plane electromagnetic wave we have instead of
Eq. (4) the following vector potential
\begin{equation}
A_\mu =\sum_{n=1}^N\sum_{s=1}^2\frac{e_{s\mu }}{\sqrt{2n\kappa
_0V}}\left[ c_{ks}^{-}e^{in(\kappa x)}+c_{ks}^{+}e^{-in(\kappa
x)}\right] .  \label{27}
\end{equation}

Taking into account Eqs. (13)-(17) and (27), equation (11) is
transformed to
\[
\biggl\{ \sum_{n,s}n\left( \frac{\partial ^2}{\partial \xi
_{ns}^2}-\xi _{ns}^2\right) +\frac 1{(q\kappa )}\left(
\sum_{n,s}\frac{b_{s\mu }}{\sqrt{n} }\xi _{ns}\right)
^2-2\sum_{n,s}\frac{\alpha _s}{\sqrt{n}}\xi _{ns}
\]
\begin{equation}
-\frac{W(q\kappa )}{e^2}\left( \sum_{n,s}b_{s\mu }\sqrt{n}\frac
\partial {\partial \xi _{ns}}\right) ^2+\frac{q^2+m^2}{(q\kappa
)}\biggr\} \chi (\xi )=0  ,\label{28}
\end{equation}
where we use the notations
\begin{equation}
b_{s\mu }=be_{s\mu } ,\hspace{0.3in}b=\frac e{\sqrt{\kappa _0V}} ,
\hspace{0.3in} \alpha _s=\frac{(qb_s)}{(q\kappa )} ,
\hspace{0.3in}W=\frac{\alpha +\beta }m ,\label{29}
\end{equation}
and the scalar products $(q\kappa )\equiv q_\mu \kappa _\mu $,
$q^2=q_\mu ^2$ and so on. Taking into consideration the equality
$b_{s\mu }b_{s^{\prime }\mu }=b^2\delta _{ss^{\prime }}$ and
introducing
\begin{equation}
M=\frac{q^2+m^2}{(q\kappa )} ,\hspace{0.3in}\delta
=\frac{b^2}{(q\kappa )} ,\hspace{0.3in}\gamma =\frac{W(q\kappa
)^2}{e^2}\delta  ,\label{30}
\end{equation}
Eq. (28) is converted into
\[
\sum_{s=1}^2\biggl\{ \sum_{n=1}^Nn\left( \frac{\partial
^2}{\partial \xi _{ns}^2}-\xi _{ns}^2\right) +\delta \left(
\sum_{n=1}^N\frac{\xi _{ns}}{\sqrt{n}}\right) ^2-2\alpha
_s\sum_{n=1}^N\frac{\xi _{ns}}{\sqrt{n}}
\]
\begin{equation}
-\gamma \left( \sum_{n=1}^N\sqrt{n}\frac \partial {\partial \xi
_{ns}}\right) ^2+M\biggr\} \chi (\xi )=0 . \label{31}
\end{equation}

Variables with different polarization index $s=1,2$ are separated
in equation (31), and, therefore, we can look for a solution to
Eq. (31) in the form $ \chi (\xi )=f_1(\xi )f_2(\xi )$, where the
function $f_s(\xi )$ obeys the equation
\[
\biggl\{ \sum_{n=1}^Nn\left( \frac{\partial ^2}{\partial \xi
_{ns}^2}-\xi _{ns}^2- \frac{2\alpha _s}{\sqrt{n}}\xi _{ns}\right)
+\delta \left( \sum_{n=1}^N\frac{ \xi _{ns}}{\sqrt{n}}\right) ^2
\]
\begin{equation}
-\gamma \left( \sum_{n=1}^N\sqrt{n}\frac \partial {\partial \xi
_{ns}}\right) ^2+\varepsilon _s\biggr\} f_s(\xi )=0  \label{32}
\end{equation}
with the condition $M=\varepsilon _1+\varepsilon _2$. Equation
(32) represents the system of interacting oscillators. To avoid
the term in Eq. (32) which is linear in $\xi _{ns}$ and to have
the unit coefficient at the second derivatives we made the linear
transformation
\begin{equation}
z_{ns}=\frac{\xi _{ns}}{\sqrt{n}}+a_{ns}  \label{33}
\end{equation}
with the constraint
\begin{equation}
n^2a_{ns}-\alpha _s-\delta \sum_{m=1}^Na_{ms}=0.  \label{34}
\end{equation}
It is easy to find the following solution to Eq. (34)
\begin{equation}
a_{ns}=\frac{\alpha _s}{n^2\left( 1-\delta \sigma _N\right) } ,
\label{35}
\end{equation}
where
\[
\sigma _N=1+\frac 1{2^2}+...+\frac 1{N^2}.
\]
In the case of infinite number of modes of quantized
electromagnetic waves ($ N\rightarrow \infty )$ $\sigma
_N\rightarrow \sigma _\infty =\pi ^2/6$. In new variables (33) Eq.
(32) takes the form
\[
\biggl\{ \sum_{n=1}^N\left( \frac{\partial ^2}{\partial
z_{ns}^2}-n^2z_{ns}^2\right) +\delta \left( \sum_nz_{ns}\right) ^2
\]
\begin{equation}
-\gamma \left( \sum_{n=1}^N\frac \partial {\partial z_{ns}}\right)
^2+\frac{ \alpha _s^2\sigma _N}{1-\delta \sigma _N}+\varepsilon
_s\biggr\} f_s(z)=0. \label{36}
\end{equation}

It is convenient to rewrite equation (36) as follows:
\[
\biggl\{ \sum_{n,m}\biggl[ \left( \delta _{nm}-\gamma \right)
\frac{\partial ^2}{\partial z_{ns}\partial z_{ms}}-\left(
n^2\delta _{nm}-\delta \right) z_{ns}z_{ms} \biggr]
\]
\begin{equation}
+\frac{\alpha _s^2\sigma _N}{1-\delta \sigma _N}+\varepsilon
_s\biggr\} f_s(z)=0.  \label{37}
\end{equation}
Eq. (37) still describes the system of interacting oscillators. To
diagonalize Eq. (37) let us introduce the matrices $A$, $B$, $C$
and $D$ in N-dimensional space:
\[
A=I-\gamma C ,\hspace{0.3in}B=D-\delta C ,
\]
\begin{equation}
C=\left(
\begin{array}{cccc}
1 & 1 & . & 1 \\
1 & 1 & . & 1 \\
. & . & . & . \\
1 & 1 & . & 1
\end{array}
\right) ,\hspace{0.3in}D=\left(
\begin{array}{cccc}
1 & 0 & . & 0 \\
0 & 2^2 & . & 0 \\
. & . & . & . \\
0 & 0 & . & N^2
\end{array}
\right)  \label{38}
\end{equation}
with the matrix elements $A_{nm}=\delta _{nm}-\gamma C_{nm}$, $
B_{nm}=D_{nm}-\delta C_{nm}$, $D_{nm}=n^2\delta _{nm}$; $I$ is the
unit matrix. Using matrices (38) Eq. (37) is converted into
equation
\begin{equation}
\biggl\{ \left( \frac \partial {\partial z}\right) ^TA\frac
\partial {\partial z}-\left( z\right) ^TBz+\frac{\alpha _s^2\sigma
_N}{1-\delta \sigma _N}+\varepsilon _s\biggr\} f_s(z)=0 ,
\label{39}
\end{equation}
where $\partial /\partial z$, $z$ are the columns, and $(\partial
/z)^T$, $ (z)^T$ are the rows of variables $\partial /\partial
z_{ns}$, $z_{ns}$ ($ n=1,2,...,N$). Two quadratic forms for
derivatives and coordinates in Eq. (39) can be diagonalized by the
linear transformation
\begin{equation}
z=Py ,\hspace{0.3in}A^{\prime }=P^{-1}A\left( P^{-1}\right) ^T ,
\hspace{0.3in} B^{\prime }=P^TBP , \label{40}
\end{equation}
where $P^{-1}P=1$ and $P^T$ is the transposed matrix.
Transformations (40) with diagonal matrices $A^{\prime }$ and
$B^{\prime }$ guarantee that Eq. (39) becomes diagonal and
variables $y_k$ will be separated. Transformations (40) mean the
transition to normal variables $y_{ns}$ which describe
non-interacting oscillators. With the help of this variables we
construct creation $\bar{c}_{ks}^+$ and annihilation
$\bar{c}_{ks}^-$ operators of non-interacting quasi-photons
\[
\bar{c}_{ks}^{-}=\frac 1{\sqrt{2}}\left( y _{ks}+\frac \partial
{\partial y _{ks}}\right) ,\hspace{0.5in}\bar{c}_{ks}^{+}=\frac
1{\sqrt{2}}\left( y _{ks}-\frac \partial {\partial y_{ks}}\right).
\]
Then Eqs. (40) are equivalent to Bogolubov's transformations (see
also [6]). It follows from Eq. (40) that the matrix $A^{-1}$
transforms under the transformation $z=Py$ like the matrix $B$:
\begin{equation}
\left( A^{-1}\right) ^{\prime }=P^TA^{-1}P .  \label{41}
\end{equation}

Thus, to find a matrix $P$ we should solve the characteristic
equation [12]
\begin{equation}
\det \left( B-\lambda ^2A^{-1}\right) =0 .  \label{42}
\end{equation}

Solutions to Eq. (42) $\lambda _n^2$ ($n=1,2,...,N$) are
eigenvalues, so that $BP_n=\lambda _n^2A^{-1}P_n$ or
$ABP_n=\lambda _n^2P_n$, where $P_n$ ($N$-dimensional columns) are
eigenvectors. At $P$-transformations matrices $ B$ and $A^{-1}$
(and $A$) can be diagonalized simultaneously so that (see [12])
\begin{equation}
\left( \frac \partial {\partial z}\right) ^TA\frac \partial
{\partial z}=\left( \frac \partial {\partial y}\right) ^T\frac
\partial {\partial y}\equiv \sum_{n=1}^N\frac{\partial
^2}{\partial y_{ns}^2} ,\hspace{0.3in} \left( z\right)
^TBz=\sum_{n=1}^N\lambda _n^2y_{ns}^2 .  \label{43}
\end{equation}

From definitions (38) we obtain the following relationships
\begin{equation}
C^2=NC ,\hspace{0.3in}A^2=I+\gamma \left( \gamma N-2\right) C ,
\hspace{0.3in} A^{-1}=I+\frac \gamma {1-\gamma N}C , \label{44}
\end{equation}
where $N$ is the number of modes. Taking into account equalities
(44) we find from Eq. (42) the characteristic equation as follows
\begin{equation}
\prod_{m=1}^N\left( m^2-\lambda ^2\right) \left[ 1-\left( \delta
+\frac{ \gamma \lambda ^2}{1-\gamma N}\right) \sum_{n=1}^N\frac
1{n^2-\lambda ^2}\right] =0 .  \label{45}
\end{equation}

It should be noted that by considering infinite number of modes, $
N\rightarrow \infty $, and as a result $\left( \delta +\gamma
\lambda ^2/(1-\gamma N)\right) \rightarrow \delta $. So, in this
case electromagnetic polarizabilities $\alpha $, $\beta $ does not
enter Eq. (45) and we come to the equation for point-like particle
(see [4]). As a result, the interaction of composite scalar
particles and the electromagnetic waves with the infinite number
of modes is similar to the interaction of scalar point-like
particles. To extract the effect of composite structure of scalar
particles we have to consider few number of modes or monochromatic
photons when the ratio $\gamma \lambda ^2/(1-\gamma N) $ is not
negligible. The formula (45) has the singularity at $\gamma N=1$ ,
and is valid when $\gamma N\neq 1$. Indeed, at $\gamma N=1$ we
have from Eq. (44) $A^2=A.$ It means that eigenvalues of the
matrix $A$ are $1$ and $0$ (because $A\left( A-1\right) =0$) and,
therefore, the matrix $A$ is a projection matrix which does not
have the inverse matrix. We skip this special case. Below we
analyze two cases: (i) $\gamma N\ll 1$, (ii) $\gamma N\gg 1$.

At $e=0$ ($\delta =0$) and $\gamma =0$ we come to non-interacting
oscillators, and the solution to Eq. (45) is $\lambda _n^2=n^2$ ($
n=1,2,...,N$). So eigenvalues $\lambda _n$ are, as usual, integer
numbers. In our case $\delta $ and $\gamma $ are small values (at
$V\rightarrow \infty $, $\delta \rightarrow 0$), and, thus, we
have for eigenvalues $ \lambda _n$ some small corrections to
integer numbers $n$. Then Eq. (45) is reduced into
\begin{equation}
\left( \delta +\frac{\gamma \lambda ^2}{1-\gamma N}\right)
\sum_{n=1}^N\frac 1{n^2-\lambda ^2}=1 .  \label{46}
\end{equation}
Eq. (46) defines the energies of quasi-photons.

Let us consider the following cases:

I. $\gamma N\ll 1$. Approximate solutions to Eq. (46) can be
looked for as the expansion $\lambda _n^2-n^2$ in the small
parameters $\delta $, $ \gamma $. We leave only terms in order of
$\delta $, $\delta ^2$, $\gamma ,$ $\gamma \delta $ and neglect
the higher powers $\delta ^3$, $\gamma ^2$ $ \gamma \delta ^2$ and
so on. As a result, we put
\begin{equation}
\lambda _n^2=n^2-a_n\delta -b_n\delta ^2-c_n\gamma -d_n\gamma
\delta . \label{47}
\end{equation}

After replacing expression (47) into Eq. (46), with the accepted
approximation, we obtain the coefficients $a_n$, $b_n$, $c_n$,
$d_n$ as follows:
\[
a_n=1 ,\hspace{0.3in}b_n=\rho _n ,\hspace{0.3in}d_n=2n^2\rho _n-1
, \hspace{0.3in} c_n=n^2 ,
\]
\begin{equation}
\rho _n=\sum_{m=1,m\neq n}^N\frac 1{m^2-n^2}
\hspace{0.3in}(n=1,2,...,N) . \label{48}
\end{equation}

The sum in Eq. (48) includes all terms besides the one with $m=n$.
Formulas (47), (48) define $N$ approximate, in the small
parameters $\delta $, $ \gamma $ eigenvalues $\lambda _n^2$ which
are the solutions of Eqs. (42), (45), (46). The parameter $\gamma
$ characterizing the internal structure of a scalar particle
(electromagnetic polarizabilities) occurring eigenvalues $\lambda
_n^2$.

Now we consider the particular cases of external quantized
electromagnetic fields:

1) Two monochromatic electromagnetic waves, N=2.

In this case $n=1,2$, and from Eq. (48) we find eigenvalues
\[
\lambda _1^2=1-\delta -\frac 13\delta ^2-\gamma +\frac 13\gamma
\delta ,
\]
\begin{equation}
\lambda _2^2=4-\delta +\frac 13\delta ^2-4\gamma +\frac{11}3\gamma
\delta . \label{49}
\end{equation}

2) Three monochromatic electromagnetic waves, N=3.

The eigenvalues $\lambda _n^2$ ($n=1,2,3$), found from Eq. (48),
are given by
\[
\lambda _1^2=1-\delta -\frac{11}{24}\delta ^2-\gamma +\frac
1{12}\gamma \delta ,
\]
\[
\lambda _2^2=4-\delta +\frac 2{15}\delta ^2-4\gamma
+\frac{31}{15}\gamma \delta ,
\]
\begin{equation}
\lambda _3^2=9-\delta +\frac{13}{40}\delta ^2-9\gamma
+\frac{137}{20}\gamma \delta .  \label{50}
\end{equation}

II. $\gamma N\gg 1$. In this case using the known sum [13]
\begin{equation}
\sum_{n=1}^\infty \frac 1{n^2-\lambda ^2}=\frac 1{2\lambda }\left(
\frac 1\lambda -\pi \cot \pi \lambda \right) , \label{51}
\end{equation}

Eq. (45) reduces to (see [4])
\begin{equation}
\left( 2\lambda ^2-\delta \right) \sin \pi \lambda +\delta \pi
\lambda \cos \pi \lambda =0 .  \label{52}
\end{equation}

With the accuracy in order of $\delta ^2$ the solution to Eq. (52)
is given by
\begin{equation}
\lambda _n=n-\frac \delta {2n}-\frac{\delta ^2}{2n^3} . \label{53}
\end{equation}

In this case, there is not an effect of composite structure of a
scalar particle and we arrive at the electromagnetic interaction
of a scalar point-like particle. Therefore, this case is not of
our interest.

Now we find the matrix $P$ which diagonalizes simultaneously
matrices $ A^{-1} $ and $B$ in accordance with formulas (40),
(41). To find the eigenvectors $P_n$ of the matrix $AB$
($ABP_n=\lambda _n^2P_n$), let us define the adjoint matrix $P$
which is the solution of the matrix equation
\begin{equation}
\left( B-\lambda ^2A^{-1}\right) P=\det \left( B-\lambda
^2A^{-1}\right) . \label{54}
\end{equation}

Matrix elements of $P$ are minors of the $\det \left( B-\lambda
^2A^{-1}\right) $ [12]. If $\lambda ^2$ are roots of the
characteristic equation (42) ($\lambda =\lambda _n$) then columns
($P_n$, $n=1,2,...,N$) of the matrix $P$ are eigenvectors, so that
$\left( B-\lambda _n^2A^{-1}\right) P_n=0$. The calculation of
minors mentioned gives matrix elements of the matrix $P$ as
follows:
\begin{equation}
P_{nm}=\frac{v_m}{n^2-m^2} , \label{55}
\end{equation}
where coefficients $v_m$ can be found by the constraints
\begin{equation}
P^TA^{-1}P=1 ,\hspace{0.3in}P^TBP=\parallel \lambda _n^2\delta
_{nm}\parallel \equiv \left(
\begin{array}{cccc}
\lambda _1^2 & 0 & . & 0 \\
0 & \lambda _2^2 & . & 0 \\
. & . & . & . \\
0 & 0 & . & \lambda _n^2
\end{array}
\right) .  \label{56}
\end{equation}

As the matrix $A^{-1}$ is positive-defined, the conditions (56)
can be satisfied [12]. The first equality $P^TA^{-1}P=1$ leads to
\begin{equation}
v_nv_k\left( \Sigma _{nk}+\frac \gamma {1-\gamma N}\Sigma _n\Sigma
_k\right) =\delta _{nk} , \label{57}
\end{equation}
\begin{equation}
\Sigma _{nk}=\sum_{m=1}^N\frac 1{\left( m^2-\lambda _n^2\right)
\left( m^2-\lambda _k^2\right) } ,\hspace{0.3in}\Sigma
_n=\sum_{m=1}^N\frac 1{m^2-\lambda _n^2} .  \label{58}
\end{equation}

It easy to verify that Eq. (57) at $n\neq k$ satisfied because in
accordance with Eq. (46) $\Sigma _n=\left( 1-\gamma N\right)
/\left[ \delta \left( 1-\gamma N\right) +\gamma \lambda
_n^2\right] $. At $n=k$ Eq. (57) allows us to find $v_n$ as
follows
\begin{equation}
v_m=\sqrt{\frac{1-\gamma N}{\left( 1-\gamma N\right) \Sigma
_{mm}+\gamma \Sigma _m^2}} .  \label{59}
\end{equation}

Practically the second term in Eq. (57) is smaller then the first
term due to the small factor $\gamma $. Therefore, at the
approximate calculations, taking into consideration small
parameters $\beta $ and $\gamma $, we can neglect the term $\gamma
\Sigma _m^2$ in Eq. (59).

The second equation in (56) is also satisfied taking into account
Eqs. (55), (59). So, the transformations (40), with the matrix $P$
(55) and coefficients (59), diagonalize the matrices $A$ and $B$
simultaneously according to Eq. (56).

After the transformations (40) with the matrix (55), and taking
into consideration Eq. (43), equation (39) becomes
\begin{equation}
\biggl\{ \sum_{n=1}^N\left( \frac{\partial ^2}{\partial
y_{ns}^2}-\lambda _n^2y_{ns}^2\right) +\frac{\alpha _s^2\sigma
_N}{1-\delta \sigma _N} +\varepsilon _s\biggr\} f_s(y)=0 .
\label{60}
\end{equation}

The variables $y_{ns}$ in Eq. (60) are separated and the finite
normalized solution at $y_{ks}\rightarrow \infty $ is given by
\begin{equation}
f_s(y)=\prod_{k=1}^Nf_{ks}(y) ,
\hspace{0.3in}f_{ks}(y)=\sqrt{\frac{\sqrt{ \lambda _k}}{\sqrt{\pi
}2^{n_k}n_k!}}H_{n_k}\left( \sqrt{\lambda _k} y_{ks}\right) \exp
\left( -\frac{\lambda _ky_{ks}^2}2\right) ,\label{61}
\end{equation}
where $H_{n_k}\left( \sqrt{\lambda _k}y_{ks}\right) $ are the
Hermite polynomials [12]. The eigenvalues $\varepsilon _s$ obey
the equality
\[
\varepsilon _s=\sum_{k=1}^N\varepsilon _{ks}-\frac{\alpha
_s^2\sigma _N}{ 1-\delta \sigma _N} ,
\]
\begin{equation}
\varepsilon _{ks}=\left( 2n_k^{(s)}+1\right) \lambda _k ,
\hspace{0.3in} n_k^{(s)}=1,2,... .  \label{62}
\end{equation}

Eq. (62) gives the energy of quasi-photons. The quantum number
$n_k^{(s)}$ is an integer and means the number of quasi-photons
with the polarization $s$ corresponding to the $k$-th mode.

\section{Momentum and mass of a particle in plane quantized electromagnetic wave}

 Using the relation $M=\varepsilon
_1+\varepsilon _2$ and Eq. (30), we find the momentum squared of
the particle-photon system
\begin{equation}
q^2=\left( q\kappa \right) \left( \varepsilon _1+\varepsilon
_2\right) -m^2 .\label{63}
\end{equation}

This is the dispersion relation for the momentum $q_\mu $ of a
particle-photon system. Let us introduce the momentum of a
particle as follows:
\begin{equation}
p_\mu =q_\mu -\frac 12\kappa _\mu \left( \varepsilon
_1+\varepsilon _2\right) .  \label{64}
\end{equation}

Using equation (63) and equality $\kappa ^2=0$ it is easy to see
that $p_\mu $ obeys the relation for the momentum of a free
particle
\begin{equation}
p^2=-m^2 .  \label{65}
\end{equation}

From Eqs. (62), (64) we come to the expression for the momentum
$q_\mu $ of a particle-photon system
\begin{equation}
q_\mu =p_\mu +\kappa _\mu \left[ \sum_{k=1}^N\lambda _k\left(
n_k^{(1)}+n_k^{(2)}+1\right) -\frac{\left( \alpha _1^2+\alpha
_2^2\right) \sigma _N}{2\left( 1-\delta \sigma _N\right) }\right]
.  \label{66}
\end{equation}

According to Eqs. (47), (48) eigenvalues $\lambda _k^2$ are given
by
\begin{equation}
\lambda _k^2=k^2-\delta -\rho _k\delta ^2-k^2\gamma -\left(
2k^2\rho _k-1\right) \gamma \delta .  \label{67}
\end{equation}

From Eq. (67) we derive approximate (at small $\delta $, $\gamma $
up to the orders $O(\delta ^2)$, $O(\gamma \delta )$) values
$\lambda _k$:
\[
\lambda _k=k-\epsilon _k,
\]
\begin{equation}
\epsilon _k=\frac 1{2k}\delta +\frac{1+4k^2\rho _k}{8k^3}\delta
^2+\frac k2\gamma +\frac{4k^2\rho _k-1}{4k}\gamma \delta .
\label{68}
\end{equation}

Let us introduce the quasi-momentum of a scalar particle (see Eq.
(26)) $P_\mu $ which is a difference between a momentum of a
particle-photon system (total momentum) $q_\mu $, a momentum of
photons in the quantized electromagnetic wave
\begin{equation}
P_\mu =q_\mu -\kappa _\mu \sum_{k=1}^Nk\left(
n_k^{(1)}+n_k^{(2)}+1\right) , \label{69}
\end{equation}
where $k_\mu =k\kappa _\mu $ is the momentum of a photon. Then
from Eqs. (29), (30), (66), (68), (69) we arrive at
\[
P^2=-m^2-b^2\sum_{k=1}^N\frac 1k\left(
n_k^{(1)}+n_k^{(2)}+1\right) -\gamma (P\kappa)\sum_{k=1}^N k\left(
n_k^{(1)}+n_k^{(2)}+1\right)
\]
\[
-b^4\sum_{k=1}^N\left[ \frac{4k^2\rho _k+1}{4(P\kappa )k^3}+\frac{
W(P\kappa )\left( 4k^2\rho _k-1\right) }{2e^2k}\right] \left(
n_k^{(1)}+n_k^{(2)}+1\right)
\]
\begin{equation}
-\frac{b^2\left[ (Pe_1)^2+(Pe_2)^2\right] \sigma _N}{( P\kappa
)-b^2\sigma _N} , \label{70}
\end{equation}
where $W=(\alpha +\beta )/m$. We took into account that according
to Eq. (69) $ (q\kappa )=(P\kappa )$. Equation (70) represents the
dispersion relation for a quasi-momentum $P_\mu$ of a scalar
composite particle. The second and third terms in the right side
of Eq. (70) contribute to the mass of a composite particle. So,
the effective mass of a scalar particle is given by
\[
m_{*}^2=m^2+\frac{e^2}{\kappa _0V}\sum_{k=1}^N\frac 1k\left(
n_k^{(1)}+n_k^{(2)}+1\right)
\]
\begin{equation}
+\frac{W(P\kappa)^2}{\kappa _0V}\sum_{k=1}^N k\left(
n_k^{(1)}+n_k^{(2)}+1\right) . \label{71}
\end{equation}

For the vacuum state $n_k^{(1)}=n_k^{(2)}=0$, and at $N\rightarrow
\infty $, we have the divergence sums in Eq. (71) which approach
to zero when the volume of the quantization $V\rightarrow \infty
$. In the classical case the average number of photons $\langle
n^{(s)}\rangle \rightarrow \infty $ and the volume $V\rightarrow
\infty $ but the ratio $\langle n\rangle /V$, where $\langle
n\rangle =\langle n^{(s)}\rangle $ remains the constant. Within
this limitation, creation and annihilation operators are replaced
by the c-numbers in accordance with relations
\begin{equation}
c_{ks}^{-}=\left( n_k^{(s)}\right) ^{1/2}\exp (i\vartheta _{ks}),
\hspace{0.3in}c_{ks}^{+}=\left( n_k^{(s)}\right) ^{1/2}\exp
(-i\vartheta _{ks}) , \label{72}
\end{equation}
where $\vartheta _{ks}$ is the phase of the classical
electromagnetic wave. Then Eq. (27) represents the Fourier
transformation of the classical electromagnetic plane wave. It is
easy to verify that Eq. (71) agrees  with Eq. (25) at classical
limit. Two last terms in equation (70) at $V\rightarrow \infty $
and $\langle n_k\rangle /V=$const trend to zero and do not
contribute to the mass in the case of classical electromagnetic
waves. These terms do not vanish only for quantized waves at the
constant volume $V$. Thus, for the classical case, the
quasi-momentum of a scalar composite particle obeys the relation
$P^2+m_{*}^2=0$. The term in Eq. (70) containing the parameter
$\gamma $ (electromagnetic polarizabilities) contributes to the
mass of a composite scalar particle. At classical limit this
additional term does not vanish.

\section{Monochromatic quantized electromagnetic wave}

Let us consider the particular case when the main contribution to
the sum of electromagnetic potential (27) comes from the definite
number $n$, such that the four-momentum of all photons is $k_\mu
=n\kappa _\mu $. Then Eq. (27) becomes
\begin{equation}
A_\mu =\sum_{s=1}^2\frac{e_{s\mu }}{\sqrt{2k_0V}}\left[
c_s^{-}e^{i(kx)}+c_s^{+}e^{-i(kx)}\right] .  \label{73}
\end{equation}
and Eq. (32) is transformed to
\[
\biggl\{ \left[ 1-W(qk)a^2\right] \frac{\partial ^2}{\partial \xi
_s^2}-\left[ 1- \frac{e^2a^2}{(qk)}\right] \xi _s^2
\]
\begin{equation}
-2e\frac{(qa_s)}{(qk)}\xi _s+\varepsilon _s\biggr\} f_s(\xi )=0 ,
\label{74}
\end{equation}
where $\chi (\xi )=f_1(\xi )f_2(\xi )$, and for convenience we use
new notations $a_{s\mu }=e_{s\mu }/\sqrt{k_0V}$, $a_{s\mu
}^2=a^2$, and eigenvalues $\varepsilon _s$ obey the following
equation:
\begin{equation}
\frac{q^2+m^2}{(qk)}=\varepsilon _1+\varepsilon _2 .  \label{75}
\end{equation}

After introducing the variables
\begin{equation}
\zeta _s=\frac{\tau (\xi _s+\sigma _s)}{c^{1/4}} ,
\hspace{0.3in}\tau ^4=1- \frac{e^2a^2}{(qk)} ,\hspace{0.3in}\sigma
_s=e\frac{(qa_s)}{\tau ^4(qk)} ,\hspace{0.3in}c=1-W(qk)a^2 ,
\label{76}
\end{equation}

Eq. (74) is rewritten as
\begin{equation}
\left( \frac{\partial ^2}{\partial \zeta _s^2}-\zeta _s^2+\nu
_s\right) f_s(\zeta )=0 , \label{77}
\end{equation}
where
\begin{equation}
\nu _s=\frac{\tau ^4\sigma _s^2+\varepsilon _s}{\tau ^2\sqrt{c}}.
\label{78}
\end{equation}

The normalized and finite at $\zeta _s\rightarrow \infty $
solution to Eq. (77) is given by
\begin{equation}
f_s(\zeta )=\frac 1{\pi ^{1/4}2^{n/2}(n!)^{1/2}}H_n\left( \zeta
_s\right) \exp \left( -\frac{\zeta _s^2}2\right)  \label{79}
\end{equation}
with the condition $\nu _s=2n_s+1$, and $n_s=1,2,...$, where $n_s$
means the number of photons with polarization $s$. This equality
leads with the help of Eqs. (75), (78) to the total squared
momentum of a particle-photon system
\begin{equation}
q^2=-m^2+2\tau ^2(qk)\left[ \sqrt{c}(n_1+n_2+1)-\frac{\tau
^2(\sigma _1^2+\sigma _2^2)}2\right] .  \label{80}
\end{equation}

From Eq. (80) we find the four-vector of a total momentum
\begin{equation}
q_\mu =p_\mu +\tau ^2k_\mu \left[
\sqrt{1-W(pk)a^2}(n_1+n_2+1)-\frac{\tau ^2(\sigma _1^2+\sigma
_2^2)}2\right] ,  \label{81}
\end{equation}
where $p_\mu ^2=-m^2$, $(pk)=(qk)$ and $p_\mu $ is a momentum of a
free scalar particle. Taking into consideration the smallness of
the parameter $ W=(\alpha +\beta )/m$, we can write
$\sqrt{1-W(pk)a^2}\simeq 1-W(pk)a^2/2.$ As a result, Eq. (81)
takes the form
\[
q_\mu =q_\mu ^{pl}-\tau ^2k_\mu \frac{W(pk)a^2}2(n_1+n_2+1),
\]
\begin{equation}
q_\mu ^{pl}=p_\mu +\tau ^2k_\mu \left[ (n_1+n_2+1)-\frac{\tau
^2(\sigma _1^2+\sigma _2^2)}2\right] .  \label{82}
\end{equation}

The value $q_\mu ^{pl}$ is the total momentum of a system of
point-like scalar particle interacting with $n_s$ ($s=1,2$)
photons. It follows from Eq. (82) that electromagnetic
polarizabilities contribute to the total momentum.

With the help of Eqs. (7), (10), (17) we write out the final
solution to Eq. (2) for composite scalar particles in the external
quantized electromagnetic wave:
\[
\varphi \left( x,\xi \right) =\left( \pi
2^{n_1+n_2}n_1!n_2!\right) ^{-1/2}\exp \left\{ i\left[ \left(
qk\right) +\frac 12\left( kx\right) \sum_{s=1}^2\left(
\frac{\partial ^2}{\partial \xi _s^2}-\xi _s^2\right) \right]
\right\}
\]
\begin{equation}
\times H_{n_1}\left( \zeta _1\right) H_{n_2}\left( \zeta _2\right)
\exp \left[ -\frac 12\left( \zeta _1^2+\zeta _2^2\right) \right] .
\label{83}
\end{equation}

\section{Coherent states}

For simplicity, let us consider the case of a liner polarization
of an electromagnetic wave. Then the polarization index $s=1$ and
solution (83) and total momentum of a system (82) become
\[
\varphi _n\left( x,\xi \right) =\left( \pi ^{1/2}2^nn!\right)
^{-1/2}\exp \left\{ i\left[ (qk)+\frac 12\left( kx\right) \left(
\frac{\partial ^2}{
\partial \xi ^2}-\xi ^2\right) \right] \right\}
\]
\begin{equation}
\times H_n\left( \zeta \right) \exp \left( -\frac 12\zeta
^2\right) ,\label{84}
\end{equation}
\begin{equation}
q_\mu =p_\mu +\tau ^2k_\mu \left[ \left( 1-\frac{W(pk)a^2}2\right)
(n+\frac 12)-\frac{\tau ^2\sigma ^2}2\right] , \label{85}
\end{equation}
where we omitted the polarization index. The variable $n$ in Eqs.
(84), (85) is the number of linearly polarized photons.

The wavefunction (84) describes a charged scalar composite
particle with the arbitrary phase of the wave interacting with $n$
external photons. Because in wavefunction (84) we have two
different arguments $\xi $ and $\zeta $ let us make the expansion
[15]
\begin{equation}
H_n\left( \zeta \right) \exp \left( -\frac 12\zeta ^2\right)
=\sum_{m=1}^\infty \beta _{nm}H_m\left( \xi \right) \exp \left(
-\frac 12\xi ^2\right) , \label{86}
\end{equation}
where the coefficients $\beta _{nm}$ can be calculated from the
relationship
\begin{equation}
\beta _{nm}=\frac 1{\sqrt{\pi }2^mm!}\int_{-\infty }^\infty
d\overline{\zeta }H_n\left( \overline{\zeta }\right) H_m\left(
\overline{\xi }\right) \exp \left( -\frac{\overline{\zeta
}^2+\overline{\xi }^2}2\right) , \label{87}
\end{equation}
where $\overline{\zeta }=\tau (\overline{\xi }+\sigma )/c^{1/4}$.
Using the property of the oscillator wavefunction [16]
\begin{equation}
\left( \xi ^2-\frac{\partial ^2}{\partial \xi ^2}\right) H_n\left(
\xi \right) \exp \left( -\frac 12\xi ^2\right) =\left( 2n+1\right)
H_n\left( \xi \right) \exp \left( -\frac 12\xi ^2\right) ,
\label{88}
\end{equation}
we derive taking into account (86) the wavefunction (84):
\begin{equation}
\varphi _n\left( x,\xi \right) =N_0\sum_{m=1}^\infty \exp \left\{
i\left[ (qk)-\frac 12\left( kx\right) \left( 2m+1\right) \right]
\right\} \beta _{nm}H_m\left( \xi \right) \exp \left( -\frac 12\xi
^2\right) , \label{89}
\end{equation}
where $N_0=\left( \pi ^{1/2}2^nn!\right) ^{-1/2}$. The result of
interacting a scalar particle with a quantized electromagnetic
field is that the solution (89) represents the superposition of
the oscillator wavefunctions with all quantum numbers $m$. For
noninteracting particle, the charge $e=0$, electromagnetic
polarizabilities $\alpha =\beta =0$ ($W=0$), and parameters take
the values $\tau =1$, $\sigma =0$, $c=1$. As a result $\zeta =\xi
$, $ \beta _{nm}=\delta _{nm}$ and wavefunction (89) has only one
term with the definite $n$. To calculate the sum in equation (89),
we use the relation [16]
\begin{equation}
\sum_{m=0}^\infty \frac{z^m}{2^mm!}H_m(x)H_m(y)=\frac
1{\sqrt{1-z^2}}\exp \left[ \frac{2xyz-(x^2+y^2)z^2}{1-z^2}\right]
.  \label{90}
\end{equation}

Replacing expression (87) into Eq. (89), and using Eq. (90), we
arrive at
\[
\varphi _n\left( x,\xi \right) =N_0\sqrt{\frac z{\pi \left(
1-z^2\right) }} \exp \left[ i(qk)-\frac{\xi
^2(z^2+1)}{2(1-z^2)}\right]
\]
\begin{equation}
\times \int_{-\infty }^\infty d\overline{\xi }H_s(\overline{\zeta
})\exp \left[ -\frac{\overline{\zeta }^2}2-\frac{\overline{\xi
}^2(z^2+1)-4\xi \overline{\xi }z}{2(1-z^2)}\right] , \label{91}
\end{equation}
where $z=\exp \left[ -i\left( kx\right) \right] $,
$\overline{\zeta }=\tau ( \overline{\xi }+\sigma )/c^{1/4}$. With
the help of the generating function [16]
\begin{equation}
\sum_{m=0}^\infty \frac{t^m}{m!}H_m(x)=\exp \left( -t^2+2tx\right)
, \label{92}
\end{equation}
we can evaluate the integral in Eq. (91), and we find (see also
[15])
\[
\varphi _n\left( x,\xi \right) =N_0\sqrt{\frac{2z\left[ 1+z^2-\tau
^2(1-z^2)/c^{1/2}\right] ^n}{\left[ 1+z^2+\tau
^2(1-z^2)/c^{1/2}\right] ^{n+1}}}H_n\left( \frac{\tau \left[
\sigma (1+z^2)+2z\xi \right] c^{1/4}}{ \left[ c(1+z^2)^2-\tau
^4(1-z^2)^2\right] ^{1/2}}\right)
\]
\begin{equation}
\times \exp \left\{ i(qk)-\frac{\xi ^2\left[ (z^2-1)c^{1/2}-\tau
^2(z^2+1)\right] -4\xi \tau ^2\sigma z-\tau ^2\sigma
^2(z^2+1)}{2\left[ (1+z^2)c^{1/2}+\tau ^2(1-z^2)\right] }\right\}
.  \label{93}
\end{equation}

Wavefunction (93) corresponds to the charged scalar composite
particle interacting with $n$ external photons possessing
four-momentum $k_\mu $.

As Eq. (2) is a linear equation, the combination of solutions (93)
with different filling numbers $n$ and phases is also the solution
of Eq. (2). The distribution of phases and quantum numbers $n$ can
be arbitrary. The very important case is the coherent wave when
the state reduces the uncertainty relations for coordinates and
momentum to a minima. The coherent state is described by the
Poisson distribution of the filling numbers $n$.

The wave packet for the Poisson distribution of the photon numbers
$n$ is given by the sum [17]
\begin{equation}
\varphi _{\langle n\rangle }(x,\xi )=\exp \left( -\frac 12\langle
n\rangle \right) \sum_{n=0}^\infty \frac 1{\sqrt{n!}}\langle
n\rangle ^{n/2}\varphi _n(x,\xi ) , \label{94}
\end{equation}
where $\langle n\rangle $ is the average photon number, and we put
the phase to be zero [15].

Replacing Eq. (93) into Eq. (94), taking into consideration Eq.
(85) and using the properties of Hermite polynomials, we arrive at
the wavefunction for coherent state with the average photon number
$\langle n\rangle $ to be present in the volume $V$ :
\[
\varphi _{\langle n\rangle }(x,\xi )=\frac{\pi
^{-1/4}c^{1/4}\sqrt{2z}}{ \sqrt{(1+z^2)\sqrt{c}+\tau
^2(1-z^2)}}\exp \biggl\{ -\frac{\langle n\rangle }2
\]
\[
+i\left[ (px)+\frac{\tau ^2}2\left( \sqrt{c}-\tau ^2\sigma
^2\right) (kx)\right] +2^{-1}\left[ \sqrt{c}(1+z^2)+\tau
^2(1-z^2)\right] ^{-1}
\]
\[
\times \biggl[ \xi ^2\left[ \sqrt{c}(z^2-1)-\tau ^2(z^2+1)\right]
-4\xi \tau ^2\sigma z-\tau ^2\sigma ^2(z^2+1)
\]
\begin{equation}
-\frac{\langle n\rangle }{z^{2\tau ^2\sqrt{c}}}\left[
\sqrt{c}(1+z^2)-\tau ^2(1-z^2)\right] +\frac{2\sqrt{2}}{z^{\tau
^2\sqrt{c}}}\sqrt{\langle n\rangle }\tau c^{1/4}\left[ \sigma
(1+z^2)+2z\xi \right] \biggl] \biggl\} . \label{95}
\end{equation}

The variables $\tau $, $a$, and $\sigma $ contain the volume of
the quantization of electromagnetic waves $V$. At the particular
case when electromagnetic polarizabilities $\alpha =\beta =0$
($W=0,$ $c=1$), we arrive at the case of pointlike scalar
particles (see [15]).

Consider the classical case when the number of photons $n$ and the
volume $V$ go to infinity but the photon density $\langle n\rangle
/V$ remains constant. The coherent states describe the wave packet
which is localized at the point $\xi _0=\sqrt{2\langle n\rangle }$
[17, 18], where wavefunction (95) does not vanish. Taking into
account that $a_\mu =e_\mu /\sqrt{k_0V}$ is a small parameter and
expanding the variables $\tau ^2\simeq 1-e^2a^2/2(pk)$,
$\sqrt{c}\simeq 1-W(pk)a^2/2$, we arrive at the limit of the
expression (95) ($\langle n\rangle /V=$const):
\[
\varphi (x)=\lim_{\langle n\rangle \rightarrow \infty }\varphi
_{\langle n\rangle }(x,\xi )=\pi ^{-1/4}\exp \biggl\{
-\frac{\left( \xi -\sqrt{2\langle n\rangle }\right) ^2}2
\]
\[
+i\biggl[ (px)+\frac{e(pa_1)}{(pk)}\sin
(kx)-\frac{e^2a_1^2}{4(pk)}\left( (kx)+\frac 12\sin 2(kx)\right)
\]
\begin{equation}
-\frac{Wa_1^2(pk)}4\left( (kx)-\frac 12\sin 2(kx)\right) \biggl]
\biggl\} ,\label{96}
\end{equation}
where $a_{1\mu }=a_\mu \sqrt{2\langle n\rangle }$. Wavefunction
(96) is the product of the oscillator eigenfunction for the ground
state [18]
\begin{equation}
\varphi _{osc}=\pi ^{-1/4}\exp \left\{ -\frac{\left( \xi
-\sqrt{2\langle n\rangle }\right) ^2}2\right\} , \label{97}
\end{equation}
and the solution to Eq. (2) in the field of the classical linearly
polarized electromagnetic wave [1]:
\[
\varphi _{cl}(x)=\exp i\biggl[ (px)+\frac{e(pa_1)}{(pk)}\sin (kx)
\]
\begin{equation}
-\frac{e^2a_1^2}{4(pk)}\left( (kx)+\frac 12\sin 2(kx)\right)
-\frac{ Wa_1^2(pk)}4\left( (kx)-\frac 12\sin 2(kx)\right) \biggl]
, \label{98}
\end{equation}
where the average energy density of the coherent electromagnetic
wave is $ k_0^2a_1^2/2=k_0\langle n\rangle /V$.

\section{Conclusion}

The exact solutions of the equation for a composite scalar
particle (pion or kaon) in the field of a plane quantized
electromagnetic wave have been obtained. For the case of the
monochromatic wave, we have studied the coherent states with the
Poisson distribution of the photon number. When the filling number
approaches to infinity with the constant density of photons, the
wavefunction is converted into the solution for a composite
particle in the classical electromagnetic wave. The solutions
obtained can be used for the investigation of the behavior of
pions in strong electromagnetic fields where nonperturbative
effects are essential. The case of quantized electromagnetic
fields is important when the photon number is not large enough and
there is an interaction of a particle with separate photons. The
found wavefunction of a composite scalar particle can be applied
for solving complex problems when the semiclassical approach does
not work.

\end{document}